\documentclass [twocolumn,
 amsymb,amsmath,nobibnotes,superscriptaddress,aps,prb]{revtex4}

 \usepackage {bm}% bold math
 \usepackage{graphicx}
 \usepackage{amsmath}

\begin{document}
 \title{In-plane noncollinear exchange coupling mediated by helical edge states in  Quantum Spin Hall system}
 \author{Jinhua Gao}
 \affiliation{Department of Physics, and Center of Theoretical and Computational Physics, The University of Hong Kong,
 Hong Kong, China}
 \author{Weiqiang Chen}
 \affiliation{Department of Physics, and Center of Theoretical and Computational Physics, The University of Hong Kong,
 Hong Kong, China}
 \author{X. C. Xie}
 \affiliation{Beijing National Lab for Condensed Matter Physics and Institute of Physics, Chinese Academy of Sciences, Beijing 100
 190, China}
 \affiliation{Department of Physics, Oklahoma State University, Stillwater, Oklahoma 74078}
 \author{Fu-chun Zhang}
 \affiliation{Department of Physics, and Center of Theoretical and Computational Physics, The University of Hong Kong,
 Hong Kong, China}

 \date{April 9 2009}
 \begin{abstract}
 We  study the Ruderman-Kittel-Kasuya-Yoshida (RKKY)
  interaction mediated by helical edge states in  quantum spin
 hall system. The helical edge states induce an in-plane  noncollinear exchange
 coupling between two local spins, in contrast to the isotropic coupling induced in
 normal metal. The angle between the two local spins in the ground state
 depends on the Fermi level. This property may be used to
control the angle of spins by tuning the electric gate.

 \end{abstract}
 \maketitle
 \section{introduction}
 Recently the study of quantum spin hall (QSH) state has attracted much
 attention\cite{ref1,ref2,ref3,ref4,ref5} for its interesting topology
 and for its potential applications in the field of spintronics.

 The well-known examples of the topologically nontrivial states are
 the integer and fractional quantum hall states, where the
 quantization of the Hall conductance is protected by a topological
 invariant. The QSH insulator is  a  novel topologically insulating phase with
 time-reversal symmetry.  A 2-dimensional QSH insulator has a charge excitation gap in the
 bulk and gapless helical edge states. Two states with opposite
 spin-polarization counter-propagate at a given edge\cite{ref1, ref4}.
Due to the time reversal symmetry, these
 spin-filtered edge states are stable against weak interaction and
 disorder\cite{ref4,ref6}. Hence, they can be viewed as effective spin and
 charge conducting channels, and could be used to construct promising spintronic devices with low
 power consumption.

 RKKY interaction is an effective interaction between two local spins
 mediated by conduction electrons\cite{ref7}. It plays an important
 role in many fields of solid state physics, e.g. giant magnetoresistance (GMR)\cite{ref8}, dilute
 magnetic semiconductor (DMS)\cite{ref9}.
 More recently, people propose that the controllable RKKY
 interaction can be used to  manipulate  the quantum states of the
 local spins, a crucial point for the spintronics and quantum computing\cite{ref10,ref11}. Thereafter,
 the RKKY interactions  in different spintronic
 materials, e.g. spin-orbital system\cite{ref12, ref13,ref14} and graphene\cite{ref15, ref16, ref17, ref18}, have been
 investigated carefully in order to facilitate the further
 development of the spintronic devices. Compared with the normal cases, the
 effective interactions between local spins in these systems exhibit rather different
 properties. The RKKY interaction in the spin-orbital system
 becomes a twisted exchange coupling since it is a spin-dependent system\cite{ref12,ref13,ref14}. The spin polarizations
 of the  two local spins are no longer collinear in this case.
 As for the graphene, due to its special electronic band dispersion, it is found that
 the RKKY interaction is ferromagnetic for local spins within
 equivalent sublattices but antiferromagnetic for opposite
 sublattices when the Fermi level is near the Dirac point\cite{ref15,ref16}.

 Being an novel spintronic material, it is an quite intriguing and
 practical problem that what the RKKY interaction in the QSH
 insulator is. Since the helical edge states are the only conducting
 channels in this system, the problem becomes that what the exchange
 interaction mediated by the helical edge states is.

 In this paper, we investigate the RKKY interaction mediated by the
 helical edge states in the QSH insulator.
 Our theoretical analyzing  is mainly based on the  simplified model
 of the helical edge state\cite{ref2,ref4,ref5}
 , in which the spin of the carriers is assumed
 to be parallel or antiparallel along the z axis as shown in Fig. 1.  Actually, it is proposed that
 the QSH insulator could be  realized in various kinds of systems, e.g. graphene \cite{ref1}, quantum
 well \cite{ref2,ref3}, or semiconductor materials with special strain gradient\cite{ref4,ref5}.
  The descriptions of the helical edge state in different systems may be
 different. However, the model we used is the simplest and most basic
 one. It grasps the primary characteristic of the helical
 states, i.e. the correlation between the spin polarization and
 propagation of the carrier, and also is the exact expression of
 the helical edge states in the
 semiconductor system with special strain gradient.
  Hence, we believe that this model is a good starting point
 for analyzing of the RKKY interaction mediated by the helical
 edge state.

 We find that the helical
 property and the linear dispersion of the edge conducting electrons will lead to  an in-plane and
 noncollinear exchange coupling between two local spins along the edge. In the asymptotic limit, this interaction
 has a simple expression. We can see that the angle between the two spins can be  controlled by adjusting the
 the Fermi energy of the system. When the Fermi surface is near the Dirac point, i.e. the
 Fermi energy is near zero, the effective interaction  becomes a
 constant antiferromagnetic exchange coupling. Actually, due to the
 helicity, pure spin current can be achieved in these edge states.
 Hence, this effective coupling can also be viewed as RKKY
 interaction mediated by pure spin current.

 This paper is organized as follows. In Sec. II, the RKKY
 interaction mediated by the helical edge states of QSH insulator is
 derived. In Sec. III, we discuss the special properties of this
 exchange interaction. Finally, a brief summary will be given in
 Sec. IV.

 \section{model and formalism}
 We consider the simplest model of helical edge states which has been successfully used in the study of the
 tunneling properties of the helical edge states in the QSH
 insulator\cite{ref4,ref19,ref20}. The schematic is shown in Fig. 1.
 The helicity
 correlates the spin polarization with the propagation. Here, we
 assume that the right(left) movers $\psi_{R \uparrow}$ ($\psi_{L \downarrow}$) carry spin up(down).  In the
 noninteracting case, the linearized Hamiltonian is
 \begin{equation}\label{h0}
 H_0=-v_F \int dx (\psi^+_{R \uparrow} i\partial_x \psi_{R
 \uparrow} - \psi^+_{L \downarrow} i \partial_x \psi_{L
 \downarrow} )
 \end{equation}
 where $v_F$ is the Fermi velocity.

\begin {figure}
 \includegraphics[width=8cm]{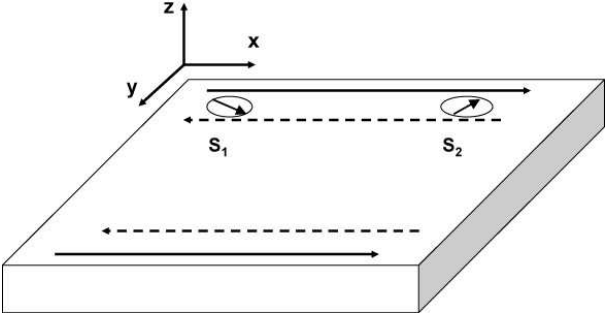}\\
 \caption{Schematic of the RKKY interaction between two local spins
 mediated by the helical edge states in QSH insulator. The QSH insulator is assumed to be a finite strip geometry
 which is
 infinite  along the x direction. Two local
 spins $S_1$ and $S_2$ are located along one edge of the QSH
 insulator. The solid (dashed) line represent the up spin right movers (down spin left movers) of the
 helical edge states on this edge.
         }
 \end  {figure}

 The localized spins are magnetic impurities and are
 denoted by $S_1$ and $S_2$. Normally the Kondo coupling between the local
 moments and conducting electrons are
 \begin{equation}\label{sd coupling}
 H_1=-\frac{J}{2}\sum_{\substack{i=1,2 \\
 \alpha,\beta = \uparrow,\downarrow}}
 \sigma_{\alpha \beta} \cdot S_i \int dx \psi^+_\alpha (x)
 \delta(x-x_i) \psi_\beta (x)
 \end{equation}
 where J is the coupling constant, $\sigma_{\alpha \beta}$ is the
 spin operator of the conducting electrons. $\alpha$ and $\beta$ are
 the spin indices. $i=1,2$ is the index of the local spins since  in order to study the RKKY interaction, we need to consider two
 local spins . For the helical edge states, the only difference
 is the spin polarization is correlated with the propagation. Hence
 in the Kondo coupling expression of the helical edge states, $\psi_{\uparrow}(x)$ ($\psi_{\downarrow}(x)$) means $\psi_{R
 \uparrow}(x)$ ($\psi_{L \downarrow}(x)$)\cite{ref4,ref21}.

 If the coupling J is small, $H_1$ can be treated as a perturbation
 on $H_0$. The RKKY interaction between two local spins $S_1$ and
 $S_2$ can be calculated from the second order perturbation
 theory\cite{ref12, ref14, ref23}
 \begin{equation}\label{rkky}
 \begin{split}
  E_{RK}=-\frac{J^2}{\pi} \texttt{Im} \int
  ^{E_F}_{-\infty}
  d \omega  Tr[(S_1 \cdot \sigma) \\ \times G^r (R_{12};\omega) (S_2 \cdot \sigma) G^r (-R_{12};\omega)]
  \end{split}
 \end{equation}
 where $E_F$ is the Fermi energy, $R_{12}=x_1 - x_2$ is the
 distance between the two local spins and Tr means the trace over
 the spin degree of freedom of conduction electrons. Therefore, the
 study of RKKY interaction has been reduced to the calculation of
 the retarded Green's function of the helical edge states.

 The definition of retarded Green's function is
 \begin{equation}
  G^r_{\alpha \beta}(xt,x't')=-i \theta(t-t') \langle\{\psi_\alpha (xt),\psi^+_{\beta}(x't')\}\rangle
 \end{equation}
 where $\alpha , \beta = \{R \uparrow , L\downarrow\}$ are the helical
 spin indices. The calculation of Green's function is quite
 straightforward. Because that there is no coupling between different spins in the noninteracting
 Hamiltonian \eqref{h0}, the nonzero Green's functions are
 $G^r_{R\uparrow R\uparrow}(x,x')$ and
 $G^r_{L\downarrow L\downarrow}(x,x')$. Take $G^r_{R\uparrow
 R\uparrow}(x,x')$ for example,
 \begin{equation}\label{green}
 \begin{aligned}
 G^r_{R\uparrow R\uparrow}(x,x';\omega) &= \int dt e^{i \omega t}
 G^r_{R\uparrow R\uparrow} (xt,x't')\\
 &= \frac{1}{2 \pi} \int dk \frac{e^{ik (x-x')} } {\omega - v_F k + i
 \eta} C_\Lambda (|k|)
  \end{aligned}
 \end{equation}
 where $\Lambda$ is the cutoff of the momentum and  $C_{\Lambda} (|k|)$ is the cutting
 off function. Since our model is a low energy approximation,
 a cutoff of the momentum is necessary and according to the
 discussion in the study of Graphene \cite{ref16}, a sharp cutoff is not suitable
 here. Therefore, we use a smooth cutting function
 \begin{equation}
 C_{\Lambda}(|k|)=e^{-\frac{|k|}{\Lambda}}.
 \end{equation}
 With this cutting off function, we get
 \begin{equation}\label{rg}
G^r_{R\uparrow R\uparrow} (x,x'; \omega)= - \frac{i}{v_F} \cdot e^{-
\frac {|\omega|} {v_F \Lambda}} \cdot e^{i \frac{\omega}{v_F}
(x-x')} \theta (x-x')
 \end{equation}
 where $\theta (x-x')$ is a step function.
 The retarded Green's function of the left movers is
 \begin{equation}\label{lg}
 G^r_{L\downarrow L\downarrow} (x,x';\omega) = \frac {i}{v_F} \cdot
 e^{- \frac{|\omega|}{v_F \Lambda}} \cdot e^{i \frac {\omega}{v_F}
 (x'-x)} \theta(x'-x)
 \end{equation}
Here, $e^{- \frac{|\omega|}{v_F \Lambda}}$ is the decay factor of
the retarded Green's function. We have to emphasize that the Green's
functions with different smooth cutting functions will have similar
form. The only difference is the decay factor.

Then substituting the Green's functions into Eq. \eqref{rkky}, we
will get the final expression of the exchange interaction between
two local spins. Without loss of generality, we set $R_{12}= x_1 -
x_2 > 0$ and then
\begin{equation}
\begin{split}
E_{RK}&=-\frac{J^2}{\pi} \textrm{Im} \int ^{E_F}_{- \infty} d\omega
[(S_1 \cdot
\sigma)_{\downarrow \uparrow}  G^r_{R \uparrow R \uparrow} (R_{12}) \\
& \times (S_2 \cdot \sigma)_{\uparrow \downarrow}
   G^r_{L \downarrow L \downarrow } (-R_{12})]  \\
  &=-\frac{J^2}{\pi} \textrm{Im} \int ^{E_F}_{- \infty} d\omega (S_{1x} S_{2x} + S_{1y} S_{2y}  + i S_{1y} S_{2x}
  \\&- i S_{1x} S_{2y}) G^r_{R \uparrow R \uparrow} (R_{12}) G^r_{L \downarrow L \downarrow } (-R_{12})
\end{split}
\end{equation}
It is clear that the range function for terms of $S_{1x} S_{2x}$ and
$S_{1y} S_{2y}$ is
\begin{equation}
\begin{split}
F_1(R_{12})=- \frac{J^2}{\pi} \textrm{Im} \int ^{E_F} _{- \infty} d
\omega G^r_{R \uparrow  R \uparrow} (R_{12}) G^r_{L \downarrow L
\downarrow} (- R_{12})
\end{split}
\end{equation}
The range function for term of $S_{1y} S_{2x}$ is

\begin{equation}
F_2(R_{12})= - \frac{J^2}{\pi} \textrm{Im} [i \int ^{E_F} _{-
\infty} d \omega  G^r_{R \uparrow  R \uparrow} (R_{12}) G^r_{L
\downarrow L \downarrow} (- R_{12})].
\end{equation}
And that for term of $S_{1x} S_{2y}$ is
\begin{equation}
\begin{split}
F_3(R_{12})  = - F_2(R_{12})
\end{split}
\end{equation}
Hence, we only need to consider the range functions $F_1(R_{12})$
and $F_2(R_{12})$.

We see that all the range functions are related to the kernel
function
\begin{equation}
K(R)= \int ^{E_F}_{-\infty} d \omega  G^r_{R \uparrow  R \uparrow}
(R) G^r_{L \downarrow L \downarrow} (- R).
\end{equation}
It depends on the Fermi energy

\begin{equation}
%\[
K(R)= \left\{ \begin{array}{ll}
                          \frac{exp[\frac{2E_F}{v_F}
(\frac{1}{\Lambda}+iR)]}{2v_F (\frac{1}{\Lambda}+iR)}  \quad E_F \leq 0  \\

\frac{1}{2v_F} \{\frac{\frac{2}{\Lambda}}{\frac{1}{\Lambda ^ 2} +
R^2}- \frac{exp[-\frac{2E_F}{v_F}(\frac{1}{\Lambda}-i
R)]}{\frac{1}{\Lambda} - i R} \}  \quad E_F>0
                    \end{array} \right.
%\]
\end{equation}

Finally, for cases $E_F \leq 0$, we get the range functions
\begin{equation}
F_1(R) = - \frac{J^2 e^{\frac{2E_F}{v_F \Lambda}}}{2 \pi v_F
(\frac{1}{\Lambda ^2} + R^2)} [\frac{sin(\frac{2E_F
R}{v_F})}{\Lambda} - R cos(\frac{2 E_F R}{v_F})]
\end{equation}
\begin{equation}
F_2(R) = - \frac{J^2 e^{\frac{2E_F}{v_F \Lambda}}}{2 \pi v_F
(\frac{1}{\Lambda ^2} + R^2)} [\frac{cos(\frac{2E_F
R}{v_F})}{\Lambda} + R sin(\frac{2E_F R}{v_F})]
\end{equation}

And for cases $E_F >0$,
\begin{equation}
\begin{split}
F_1(R)&=-\frac{J^2 e^{\frac{2(-E_F)}{v_F \Lambda}}}{2 \pi v_f
(\frac{1}{\Lambda ^2} + R^2)}\{\frac{sin[\frac{2(-E_F)
R}{v_F}]}{\Lambda}  \\& - R cos[\frac{2 (-E_F) R}{v_F}]\}
\end{split}
\end{equation}
\begin{equation}
\begin{split}
F_{2}(R)&=-\frac{J^2}{2 \pi v_F (\frac{1}{\Lambda^2}+R^2)} \{
\frac{2}{\Lambda} - e^{-\frac{2E_F}{v_F \Lambda}}
\\ & \times [\frac{cos(\frac{2 E_F R }{v_F})}{\Lambda} - R sin(\frac{2 E_F
R}{v_F})]\}
\end{split}
\end{equation}

\section{Discussion}
In the continuum limit $\Lambda |x-x'|=\infty$, we will get a
rather simple expression
\begin{equation}\label{cr}
\begin{split}
E_{RK}= \frac{J^2}{2 \pi v_F |R|} [cos\alpha \cdot (S_{1x}
S_{2x}+S_{1y} S_{2y}) \\ - sin\alpha \cdot (S_{1y} S_{2x} - S_{1x}
S_{2y})]
\end{split}
\end{equation}
where $R=x-x'$ is the distance between the two local spins and
$\alpha = \frac{2 |R| E_F}{v_F}$.

If we consider the two local spins $S_1$ and $S_2$ as classical
spins, Eq.\eqref{cr} can be transformed into
\begin{equation}
E_{RK}=\frac{J^2 M^2}{2 \pi v_F |R|} \cdot sin \theta_1 sin \theta_2
\cdot cos[(\phi_1 - \phi_2)+ \alpha]
\end{equation}
Here, $(M,\theta_{1}, \phi_{1})$ and $(M, \theta_2, \phi_2) $ are
the spherical coordinates of the spin vector $S_1$ and $S_2$. We can
see that only if $\theta_1 = \theta_2 = \pi /2$ and $cos[(\phi_1 -
\phi_2) + \alpha]=-1$, the system will have its lowest energy.  In
this case, $\theta_1 = \theta_2 = \pi /2$ means that $S_1$ and $S_2$
are in-plane. And $cos[(\phi_1 - \phi_2) + \alpha] = -1$ shows that
the exchange interaction is noncollinear. It means that the
effective exchange interaction mediated by the helical edge states
is an in-plane and noncollinear coupling. Actually, it is easy to
see from the expressions  that in-plane and noncollinear are the
general characteristics of the helical edge states mediated exchange
interaction which does not depend on the continuum limit.

In the continuum limit, the angle between the local spins is
determined by $\alpha=\frac{2|R|E_F}{v_F}$, i.e. the Fermi energy
$E_F$ and the distance $R$. Especially when the Fermi energy $E_F =
0$, i.e. the Fermi level is around the Dirac point, $\alpha = 0$ and
$\phi_1 - \phi_2 = \pi$. It means that the exchange coupling becomes
a constant and is always antiferromagnetic. Here, the Dirac point is
the crossing of the bands of right and left movers. However, in
general cases, we do not have a simple formula about the angle
between the local spins. It should be determined though concrete
numerical calculation.

The special characteristics of the helical edge states mediated
exchange interaction  result from the interplay between the helicity
and the linear band dispersion. As shown in former studies, without
helicity, if the system is spin-independent, the coupling will have
similar form $F(R)S_1 \cdot S_2$ and the only difference is just the
range function $F(R)$. In our case, the helicity makes the system
spin-dependent: though the matrix of Green's function is still
diagonal but $G^r_{R\uparrow R\uparrow}(x,x';\omega) \neq G^r_{L
\downarrow L\downarrow} (x,x';\omega)$. It is the main reason of the
noncollinear behavior of the exchange coupling. In addition to the
helicity, the linear band dispersion induces opposite  step
functions into the Green's functions as shown in Eq. \eqref{rg} and
\eqref{lg}. Actually, the in-plane characteristic of the exchange
interaction mainly results from these step functions.

\section{conclusion}
In summary,  based on the simplified model of the helical edge
states, we have investigated the helical edge states that mediate
RKKY interaction between local spins in the QSH system. Since the
conducting electrons in the helical edge states are the only
carriers in the QSH insulator, this exchange interaction is probably
the only possible mechanism of the exchange coupling between local
spins in such a promising spintronic system. Furthermore, due to the
helicity, i.e. the correlation between the spin polarization and
propagation, it is believed that pure spin current can be realized
in this edge state. It means that this exchange interaction is
actually an exchange interaction mediated by pure spin current.
Hence, this problem is not only of fundamental interest but also
useful for the future development of spin-based devices in such
systems.

We analyze the simplest theoretical model of the helical edge states
and concentrate on the effects of the helicity and its linear band
dispersion. We find that the RKKY interaction mediated by this
helical edge state in such a system is in-plane and noncollinear,
which is extremely different from the exchange interaction in other
systems. The angle between the local spins depends on the Fermi
energy of the system. Therefor, this effective interaction offers a
possible way to control such angle through adjusting the fermi level
by a gate. In the continuum limit, a simple expression of this
exchange interaction can be achieved. Especially, when the Fermi
level is around the Dirac point, the exchange coupling becomes a
constant antiferromagnetic one. We also point out that these
peculiar properties result from the interplay between the helicity
and the linear band dispersion.

However, our study is based on the noninteracting low energy
approximation model of the helical edge states. For a concrete
material or experimental setup, many more practical factors may need
to be included. We believe that our analysis is a good starting
point for the further investigation.

\textbf{Acknowledgments:} We  wish to acknowledge the partial
support from RGC grant of HKSAR. XCX is supported by DOE and C-Spin
center of Oklahoma.

 \end{document}